\documentclass[12pt]{article}
\usepackage{amssymb,epsfig}
\begin{document}
\baselineskip 24pt

\begin{titlepage}
\begin{center}
{\LARGE \bf Light-cone QCD sum rule approach for the $\Xi$ baryon electromagnetic form factors and the semileptonic decay $\Xi_c\rightarrow \Xi e^+\nu_e$} \vspace{1cm}

{Yong-Lu Liu\footnote{E-mil: yongluliu@nudt.edu.cn} } and {Ming-Qiu
Huang}
\\[0.5cm]
\vspace*{0.1cm} {\it Department of Physics, National University of
Defense Technology, Hunan 410073, China}

\vspace{0.6cm}
\bigskip

\vskip1.2cm

{\bf Abstract \\[10pt]} \parbox[t]{\textwidth}
{\baselineskip  24pt The electromagnetic form factors of the $\Xi$ baryons and the semileptonic decay process $\Xi_c\rightarrow \Xi e^+\nu_e$ are investigated in the frame work of the light-cone QCD sum rule method with Ioffe-type interpolating currents. Our estimates on the magnetic moments are $\mu_{\Xi^0}=-(1.75\pm0.21)\mu_N$ and $\mu_{\Xi^-}=-(1.01\pm0.11)\mu_N$. The decay width of the semileptonic process is expected to be $\Gamma(\Xi_c\rightarrow \Xi e^+\nu_e)=(6.17^{+2.24}_{-2.48})\times10^{-14}\;\mbox{GeV}$. The results make sure that the adoption of this type interpolating current improve the calculations of the magnetic form factors and give more reliable prediction for the analysis of the semileptonic decay process.}
\end{center}
\vspace{3cm}


\vspace{2mm}


\vspace{1cm}

\end{titlepage}

\newpage
\section{Introduction}

\label{sec1}
The interpolating current for a hadron is one of the most important ingredients in the nonperturbative quantum field theory, especially in the QCD sum rule method. It has been noticed that the current interpolating the baryon state is not unique early almost at the beginning when the QCD sum rules were used to the baryon \cite{Chung,Ioffe}. There were many works devoted to this issue \cite{Lee,WDW,IIG,Leinweber}, particularly for the investigation on the electromagnetic form factors of the nucleon \cite{Lenz,Aliev}, which have shown that the choice of the interpolating current may affect the result to some extent. In the previous work \cite{current}, we examine the influence due to the choice of the interpolating current for the electromagnetic form factors of the $\Lambda$ and $\Sigma$ baryons. The results tell that the usage of different interpolating currents for the baryons may give information from various aspects. Another work \cite{lamc} has shown that Ioffe-type current may give more reliable estimates for the semileptonic decay mode $\Lambda_c\rightarrow \Lambda l^+\nu_l$. Therefore, it is expected that the adoption of the Ioffe-type interpolating current may tell us more information on the same process.

In the early work \cite{xidas}, we gave the distribution amplitudes of the $\Xi$ baryon and examine their applications in the light-cone sum rule approach making use of Chernyak-Zhitnitsky (CZ-type) interpolating current. This manuscript is an extension and a complement of that investigations. Since it is difficult to measure properties of the unstable baryons, there are no accurate experiment data available for the electromagnetic form factors and the semileptonic decay process. Therefore, we expect the usage of the Ioffe-type interpolating current may give us more reliable predicts on these issues theoretically.

We proceed as follows. Section \ref{sec2} is devoted to derive the light cone sum rules of the $\Xi$ baryon electromagnetic form factors and estimate the magnetic moments. Section \ref{sec3} is the investigation on the semileptonic decay process $\Xi_c\rightarrow \Xi e^+\nu_e$ and the numerical analysis. Summary and conclusion are given at the end of this part.

\section{Electromagnetic form factors of $\Xi$ baryons}\label{sec2}
Electromagnetic form factors of the hadron, which reveal the internal structure of the composite particle, are elementary parameters to be investigated both theoretically and experimentally. The natural way to investigate the electromagnetic form factors is to measure the scattering of the electron off the baryon. However, this is rather difficult for short-lived hadrons. In practice, meson photoproduction and scattering off a baryon are major tools to gain a deep insight into the dynamics of baryons. The study on $N\rightarrow \Delta$ transition form factors both experimentally and theoretically leads to a good understanding of the $\Delta$ resonance (see \cite{Pscalutsa} and references therein for a review). This may imply that kaon photoproduction of a strange baryon may give internal information of the $\Xi$ baryons, provided that there is a stable source of strange baryons, such as $\Lambda$ or $\Sigma$, which may be produced through the $N\pi$ scattering at high momentum transfer. All in all, direct measurement on this short-lifetime hyperon is difficult, so we expect theoretical analysis may provide useful information. Theoretical study on the electromagnetic process can be reduced to the related electromagnetic form factors. There have been some works on the octet baryon magnetic form factors with various models \cite{Kubis,Kim,Cauteren}, which give the properties of the physical values in different region. However, the light cone QCD sum rule approach has the advantage to provide properties of the physical values at the moderately large momentum transfer. For this reason, we will investigate the electromagnetic form factors with this method in the paper.

The electromagnetic form factors of a baryon can be defined by the matrix element of the electromagnetic current between the baryon states, which is parameterized by the so-called Dirac $F_1(Q^2)$ and Pauli $F_2(Q^2)$ form factors:
\begin{equation}
\langle B(P',s')|j_\mu^{em}(0)|B(P,s)\rangle=\bar B(P',s')[\gamma_\mu F_1(Q^2)-i\frac{\sigma_{\mu\nu} q^\nu}{2M}F_2(Q^2)]B(P,s),\label{form}
\end{equation}
where $B(P,s)$ denotes the baryon spinor with the momentum $P$ and the spin $s$, $M$ is the baryon mass, $Q^2=-q^2=-(P-P')^2$ is the squared momentum transfer, and $j_\mu^{em}$ is the electromagnetic current relevant to the baryon.

Experimentally speaking, the Dirac and Pauli form factors can be redefined by the electric and magnetic form factors $G_E(Q^2)$ and $G_M(Q^2)$:
\begin{eqnarray}
G_M(Q^2)&=&F_1(Q^2)+F_2(Q^2),\nonumber\\
G_E(Q^2)&=&F_1(Q^2)-\frac{Q^2}{4M^2}F_2(Q^2).
\end{eqnarray}
In the Breit frame, $G_E(Q^2)$ describes the distribution of the electric charge and $G_M(Q^2)$ corresponds to the magnetic current distribution. Particularly, the normalization of the electric and magnetic form factors at $Q^2=0$ is given by the electric charge $G_E(0)=e_B$ and the magnetic moment $G_M(Q^2)=\mu_B$, respectively.

\subsection{Light cone QCD sum rules for the form factors}
The form factors defined above can be determined in the moderate momentum transfer by the light-cone QCD sum rule method with the aid of the distribution amplitudes presented in Ref. \cite{xidas}. In order to give the sum rules of the form factors, we begin with the following correlation function:
\begin{equation}
T_\mu(P,q)=i \int d^4xe^{iq\cdot x}\langle
0|T\{j_{\Xi}(0)j_\mu^{em}(x)\}|{\Xi^0}(P)\rangle,\label{correlator1}
\end{equation}
in which the interpolating current of the $\Xi$ baryon field is chosen as the Ioffe-type one:
\begin{equation}
{j_{\Xi}}(x)=\epsilon_{ijk}[s^i(x)C\gamma_5 \gamma_\mu s^j(x)]\gamma_\mu q^k(x),
\end{equation}
where $q$ stands for $u$ or $d$ quark, and $i$, $j$, $k$ refer to color indices. The coupling of this type current with the baryon state is defined as
\begin{equation}
\langle 0|j_\Xi|\Xi\rangle=\lambda_1M\Xi(P).\label{norm}
\end{equation}

In the following analysis, we take into account $\Xi^0$ as an example, and the corresponding part for $\Xi^-$ is similar. By inserting a complete set of intermediate states with the same quantum numbers as those of $\Xi^0$, we get the hadronic representation of the correlation function (\ref{correlator1}) with the Eqs. (\ref{form}) and (\ref{norm}):
\begin{eqnarray}
z^\mu T_\mu(P,q)&=&\frac{\lambda_1M}{M^2-P'^2}[2P\cdot zF_1(Q^2)+\frac{P\cdot z}{M}\!\not\!q_\perp
F_2(Q^2)\nonumber\\
&&+(F_1(Q^2)+F_2(Q^2))\!\not\!z\!\not\!q+\frac{q^2}{2M}\!\not\!zF_2(Q^2)]\Xi(P)+...,\label{hadrep}
\end{eqnarray}
in which $P'=P-q$, $\!\not\!q_\perp=\!\not\!q-\frac{pq}{pz} \!\not\!z$, and the dots stand for the higher resonance and the continuum contributions. Different from the case where the CZ-type current is used \cite{xidas}, there are four kinds of Lorentz structures in the presentation. It is the same as that discussed in Refs. \cite{current,lamc} that, in principle, each Lorentz structure can give information on the electromagnetic form factors. We take into account terms proportional to $1$ and $\!\not\! q_\perp$ in the calculation for convenience.

On the other hand, the theoretical representation can be obtained with the help of the $\Xi$ baryon distribution amplitudes given in Ref. \cite{xidas}. Following the standard procedure of the light cone QCD sum rule approach, we obtain the following Borel transformed sum rules by matching the two representations:
\begin{eqnarray}
\lambda_1F_1(Q^2)&=&e_u\Big\{\int_{\alpha_{30}}^1d\alpha_3e^{-(s-M^2)/M_B^2}\{-B_0(\alpha_3)-(\frac{1}{\alpha_3}
-\frac{Q^2}{\alpha_3^2M_B^2})B_1(\alpha_3)+\frac{M^2}{M_B^2}B_2(\alpha_3)\}\nonumber\\
&&+e^{-(s_0-M^2)/M_B^2}\frac{1}{\alpha_{30}^2M^2+Q^2}\{Q^2B_1(\alpha_{30})+\alpha_{30}^2M^2B_2(\alpha_{30})\}\Big\}\nonumber\\
&&+e_s\Big\{\int_{\alpha_{20}}^1d\alpha_2e^{-(s_1-M^2)/M_B^2}\{2C_0(\alpha_2)-3C_0'(\alpha_2)-2D_0(\alpha_2)-2\frac{m_s}{\alpha_2M}E_0(\alpha_2)\nonumber\\
&&+\frac{1}{\alpha_2}[C_1(\alpha_2)-D_1(\alpha_2)]+\frac{1}{M_B^2}[\frac{2m_s^2+Q^2}{\alpha_2^2}C_1(\alpha_2)-2M^2C_2(\alpha_2)\nonumber\\
&&+M^2C_3(\alpha_2)-\frac{2M^2}{\alpha_2}C_4(\alpha_2)-M^2D_2(\alpha_2)-\frac{2Mm_s}{\alpha_2}[E_1(\alpha_2)-E_2(\alpha_2)\nonumber\\
&&+E_3(\alpha_2)]+2\frac{Mm_s}{M_B^4}[-\frac{Mm_s}{\alpha_2^2}C_4(\alpha_2)+\frac{Q^2}{\alpha_2^3}E_3(\alpha_2)+\frac{M^2}{\alpha_2}E_4(\alpha_2)]\}\nonumber\\
&&+e^{-(s_0-M^2)/M_B^2}\frac{1}{\alpha_{20}^2M^2+Q^2+m_s^2}\{(2m_s^2+Q^2)C_1(\alpha_{20})+\alpha_{20}^2M^2[-2C_2(\alpha_{20})\nonumber\\
&&+C_3(\alpha_{20})-D_2(\alpha_{20})]-2\alpha_{20}M^2C_4(\alpha_{20})-2\alpha_{20}Mm_s[E_1(\alpha_{20})-E_2(\alpha_{20})\nonumber\\
&&+E_3(\alpha_{20})]-2\frac{M^2m_s^2}{M_B^2}C_4(\alpha_{20})+\frac{2Mm_sQ^2}{\alpha_{20}M_B^2}E_3(\alpha_{20})+\frac{2\alpha_{20}M^3m_s}{M_B^2}E_4(\alpha_{20})\nonumber\\
&&+2\alpha_{20}^2\frac{d}{d\alpha_{20}}[M^2m_s^2C_4(\alpha_{20})-\frac{Mm_sQ^2}{\alpha_{20}}E_3(\alpha_{20})-M^3m_s\alpha_{20}E_4(\alpha_{20})]\nonumber\\
&&\times\frac{1}{\alpha_{20}^2M^2+Q^2+m_s^2}\}\Big\}
\end{eqnarray}
and
\begin{eqnarray}
\lambda_1F_2(Q^2)&=&2e_u\Big\{\int_{\alpha_{30}}^1d\alpha_3\frac{1}{\alpha_3}e^{-(s-M^2)/M_B^2}\{B_0'(\alpha_3)+\frac{M^2}{M_B^2}[B_1-B_2](\alpha_3)\}\nonumber\\
&&+e^{-(s_0-M^2)/M_B^2}\frac{\alpha_{30}M^2}{\alpha_{30}^2M^2+Q^2}[B_1-B_2](\alpha_{30})\Big\}\nonumber\\
&&+2e_s\Big\{\int_{\alpha_20}^1d\alpha_2e^{-(s_1-M^2)/M_B^2}\frac{1}{\alpha_2}\{-[C_0+D_0'](\alpha_2)+\frac{M^2}{M_B^2}[C_1+2C_2-C_3\nonumber\\
&&+D_1+D_2](\alpha_2)+\frac{Mm_s}{\alpha_2M_B^2}[E_1+E_5](\alpha_2)+\frac{2M^3m_s}{\alpha_2M_B^4}[E_3-E_4](\alpha_2)\}\nonumber\\
&&+2e^{-(s_0-M^2)/M_B^2}\frac{\alpha_{20}M^2}{\alpha_{20}^2M^2+Q^2+m_s^2}\{[C_1+2C_2-C_3+D_1+D_2](\alpha_{20})\nonumber\\
&&+\frac{m_s}{\alpha_{20}M}[E_1+E_5](\alpha_{20})+2\frac{Mm_s}{\alpha_{20}M_B^2}[E_3-E_4](\alpha_{20})\nonumber\\
&&-2\alpha_{20}\frac{d}{\alpha_{20}}\frac{Mm_s}{\alpha_{20}^2+Q^2+m_s^2}[E_3-E_4](\alpha_{20})\}\Big\},
\end{eqnarray}
where $s=(1-\alpha_3)M^2+\frac{1-\alpha_3}{\alpha_3}Q^2$, $s_1=(1-\alpha_2)M^2+\frac{1-\alpha_2}{\alpha_2}Q^2+\frac{m_s^2}{\alpha_2}$, and
\begin{eqnarray}
B_0(\alpha_3)&=&\int_0^{1-\alpha_3}d\alpha_1V_3(\alpha_1,1-\alpha_1-\alpha_3),\nonumber\\
B_0'(\alpha_3)&=&\int_0^{1-\alpha_3}d\alpha_1V_1(\alpha_1,1-\alpha_1-\alpha_3),\nonumber\\
B_1(\alpha_3)&=&(\widetilde V_1-\widetilde V_2-\widetilde V_3)(\alpha_3),\nonumber\\
B_2(\alpha_3)&=&(\widetilde V_4-\widetilde V_3)(\alpha_3),\nonumber\\
C_0(\alpha_2)&=&\int_0^{1-\alpha_2}d\alpha_1V_1(\alpha_1,\alpha_2,1-\alpha_1-\alpha_2),\nonumber\\
C_0'(\alpha_2)&=&\int_0^{1-\alpha_2}d\alpha_1V_3(\alpha_1,\alpha_2,1-\alpha_1-\alpha_2),\nonumber\\
C_1(\alpha_2)&=&(\widetilde V_1-\widetilde V_2-\widetilde V_3)(\alpha_2),\nonumber\\
C_2(\alpha_2)&=&(-\widetilde V_1+\widetilde V_3+\widetilde V_5)(\alpha_2),\nonumber\\
C_3(\alpha_2)&=&(\widetilde V_4-\widetilde V_3)(\alpha_2),\nonumber\\
C_4(\alpha_2)&=&(-\widetilde{\widetilde V}_1+\widetilde{\widetilde V}_2+\widetilde{\widetilde V}_3+\widetilde{\widetilde V}_4+\widetilde{\widetilde V}_5-\widetilde{\widetilde V}_6)(\alpha_2),\nonumber\\
D_0(\alpha_2)&=&\int_0^{1-\alpha_2}d\alpha_1A_3(\alpha_1,\alpha_2,1-\alpha_1-\alpha_2),\nonumber\\
D_0'(\alpha_2)&=&\int_0^{1-\alpha_2}d\alpha_1A_1(\alpha_1,\alpha_2,1-\alpha_1-\alpha_2),\nonumber\\
D_1(\alpha_2)&=&(-\widetilde A_1+\widetilde A_2-\widetilde A_3)(\alpha_2),\nonumber\\
D_2(\alpha_2)&=&(\widetilde A_3-\widetilde A_4)(\alpha_2),\nonumber\\
E_0(\alpha_2)&=&\int_0^{1-\alpha_2}d\alpha_1T_1(\alpha_1,\alpha_2,1-\alpha_1-\alpha_2),\nonumber\\
E_1(\alpha_2)&=&(\widetilde T_1-\widetilde T_2-2\widetilde T_7)(\alpha_2),\nonumber\\
E_2(\alpha_2)&=&(-\widetilde T_1+\widetilde T_2+2\widetilde T_8)(\alpha_2),\nonumber\\
E_3(\alpha_2)&=&(\widetilde{\widetilde T}_2-\widetilde{\widetilde T}_3-\widetilde{\widetilde T}_4+\widetilde{\widetilde T}_5+\widetilde{\widetilde T}_7+\widetilde{\widetilde T}_8)(\alpha_2),\nonumber\\
E_4(\alpha_2)&=&(-\widetilde{\widetilde T}_2+\widetilde{\widetilde T}_2+\widetilde{\widetilde T}_5-\widetilde{\widetilde T}_6+2\widetilde{\widetilde T}_7+2\widetilde{\widetilde T}_8)(\alpha_2),\nonumber\\
E_5(\alpha_2)&=&(\widetilde T_1+\widetilde T_2-2\widetilde T_3)(\alpha_2),
\end{eqnarray}
in which
\begin{eqnarray}
\widetilde F_i(\alpha_2)&=&\int_0^{\alpha_2}d{\alpha_2'}\int_0^{1-\alpha_2'}d\alpha_1F_i(\alpha_1,\alpha_2',1-\alpha_1-\alpha_2'),\nonumber\\
\widetilde{\widetilde F_i}(\alpha_2)&=&\int_0^{\alpha_2}d{\alpha_2'}\int_0^{\alpha_2'}d{\alpha_2''}\int_0^{1-\alpha_2''}d\alpha_1F_i(\alpha_1,\alpha_2'',1-\alpha_1-\alpha_2''),\nonumber\\
\widetilde F_i(\alpha_3)&=&\int_0^{\alpha_3}d{\alpha_3'}\int_0^{1-\alpha_3'}d\alpha_1F_i(\alpha_1,1-\alpha_1-\alpha_3',\alpha_3'),\nonumber\\
\widetilde{\widetilde
F_i}(\alpha_3)&=&\int_0^{\alpha_3}d{\alpha_3'}\int_0^{\alpha_3'}d{\alpha_3''}\int_0^{1-\alpha_3''}d\alpha_1F_i(\alpha_1,1-\alpha_1-\alpha_3'',\alpha_3'').
\end{eqnarray}
It is noted that in the sum rules, $V_i$, $A_i$, and $T_i$ structures can all contribute to the results, which is quite different from the CZ-type interpolating current case.

\subsection{Numerical analysis}
In the numerical analysis, the necessary input parameters are used as follows. The masses of the baryons are the centra values from the Particle Data Group (PDG) \cite{PDG}: $M_{\Xi^0}=1.315\;\mbox{GeV}$ and $M_{\Xi^-}=1.322\;\mbox{GeV}$. The strange quark mass is adopted as $m_s=0.15\;\mbox{GeV}$. The continuum threshold $s_0$ is set to be $s_0=2.8-3.0\;\mbox{GeV}^{2}$.

Another parameter which needs to be determined is the Borel parameter. As is known, there should be a working window in which the sum rules vary mildly with the Borel parameter. Our calculations show that the proper region can be chosen as $2\; \mbox{GeV}^2\leq M_B^2 \leq 4\; \mbox{GeV}^2$. In the following analysis, we take $M_B^2=3\,\mbox{GeV}^2$.

\begin{figure}
\begin{minipage}{7cm}
\epsfxsize=6cm \centerline{\epsffile{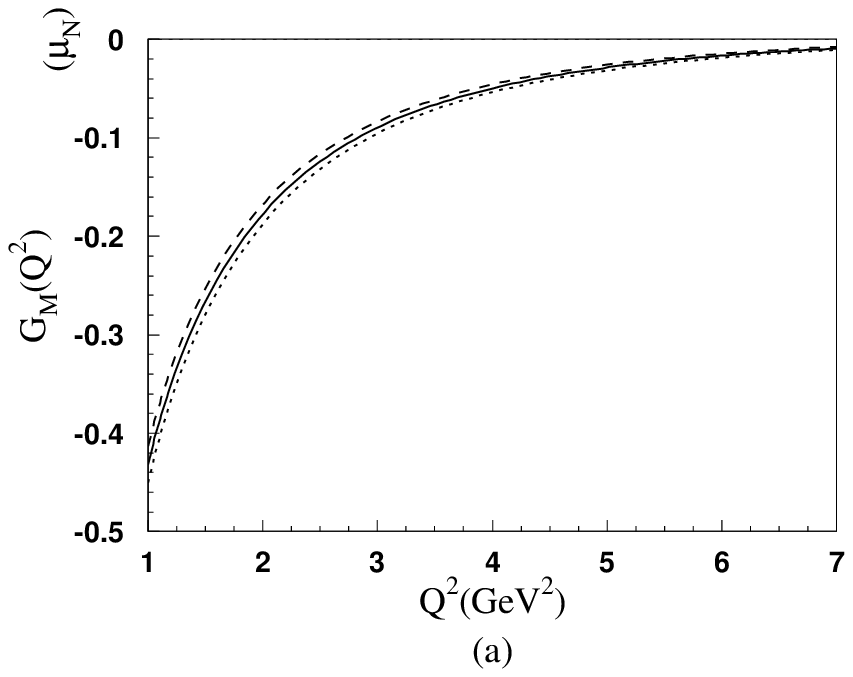}}
\end{minipage}
\hfill
\begin{minipage}{7cm}
\epsfxsize=6cm \centerline{\epsffile{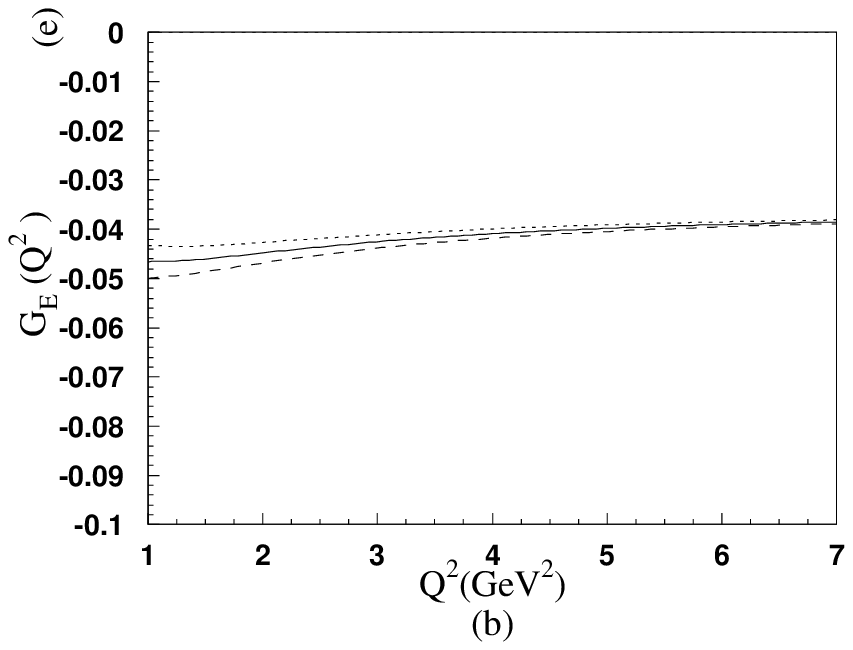}}
\end{minipage}
\caption{}\label{fig1}
\end{figure}

\begin{figure}
\begin{minipage}{7cm}
\epsfxsize=6cm \centerline{\epsffile{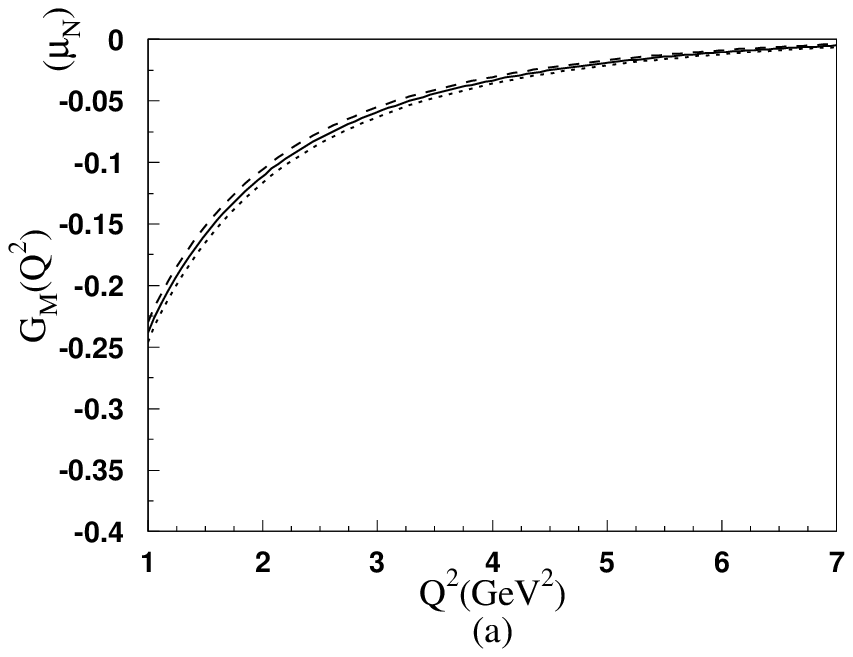}}
\end{minipage}
\hfill
\begin{minipage}{7cm}
\epsfxsize=6cm \centerline{\epsffile{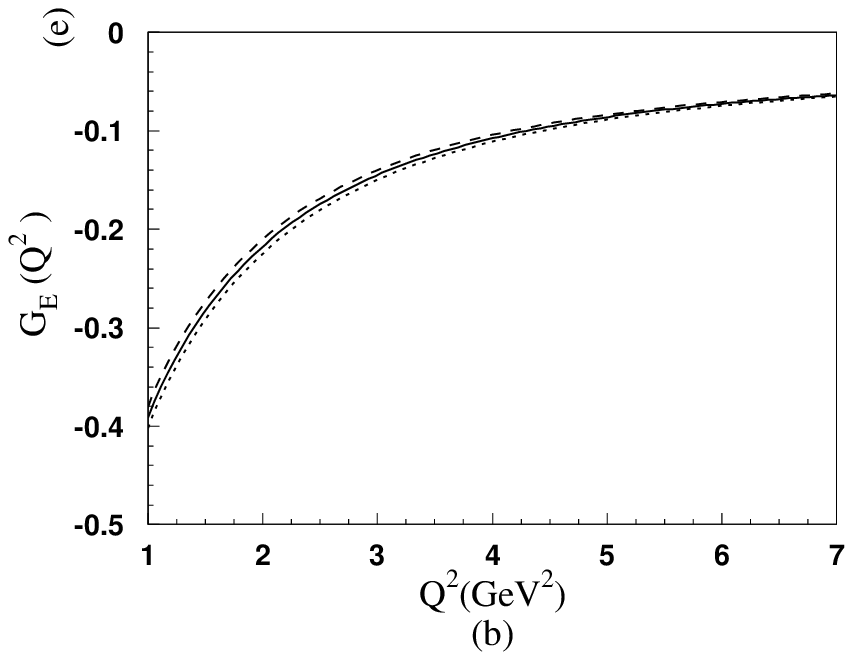}}
\end{minipage}
\caption{}\label{fig2}
\end{figure}

With the chosen parameters, we first show the dependence of the electric and magnetic form factors on the momentum transfer for $\Xi^0$ ($\Xi^-$) in Fig. \ref{fig1} (Fig. \ref{fig2}). The counterpart of the form factors in the experiment can be investigated by the elastic scattering of leptons to $\Xi$, provided that there is a source of $\Xi$ baryons available. However, the experiment condition is unavailable nowadays, so the parameters characterizing the form factors at specific point of $Q^2$, such as the magnetic moment and the mean-square radius, can be measured for simple comparisons. Therefore, it can be said that the theoretical predictions are important ways to understand the internal structure of the baryons at present.

To verify our calculations and estimate the magnetic moments, we fit the magnetic form factors by the dipole formula:
\begin{equation}\label{dipole}
\frac{1}\mu G_M(Q^2)=\frac{1}{(1+Q^2/m_0^2)^2}=G_D(Q^2),
\end{equation}
where $\mu$ is the magnetic moment of the baryon and the parameter $m_0^2$ needs to be determined experimentally. For a comparison with the results form CZ-type current, we use the values estimated from the case of CZ-type current \cite{xidas} in the regions: $m_0^2=(0.94\pm 0.05)\,\mbox{GeV}^2$ for $\Xi^0$ and $m_0^2=(0.96\pm 0.05)\,\mbox{GeV}^2$ for $\Xi^-$. Fitting the magnetic form factors by the dipole formula (\ref{dipole}), we get our estimates for the $\Xi$ baryon magnetic moments: $\mu_{\Xi^0}=-(1.75\pm0.21)\mu_N$ and $\mu_{\Xi^-}=-(1.01\pm0.11)\mu_N$. For a comparison, we provide in Table \ref{table1} the magnetic moments estimated from both the CZ-type \cite{xidas} and the Ioffe-type interpolating currents. It can be seen that the estimates on the magnetic moments with Ioffe-type current are more reliable than that from CZ-type current. However, both kinds of interpolating current give larger numerical estimates in comparison with the experiments. This is partly due to the fact that more precise estimates with dipole formula approach rely on the parameter $m_0^2$, which needs to be measured experimentally. After all, the thorough understanding of this issue needs more accurate information on the baryon distribution amplitudes with higher conformal spin contributions.

\section{The semileptonic decay process $\Xi_c\rightarrow \Xi e^+\nu_e$}\label{sec3}
Semileptonic decay processes of the charm hadrons are important channels to be studied for their importance in investigating the Cabibbo-Kobiyash-Maskawa matrix elements and testing the standard model. In our approach, in order to deal with the nonperturbative strong interaction in the hadron bound state, we separate the hadronic part from the leptonic one in the calculation. The leptonic part can be calculated with the normal method of the quantum field theory, while the hadronic one is parameterized as the weak transition form factors, which can be calculated in the light-cone QCD sum rule approach. We have studied the process $\Xi_c\rightarrow \Xi e^+\nu_e$ in the previous work \cite{xidas} with the CZ-type interpolating current. The results are somewhat larger than the experimental data. In the present paper, we adopt the Ioffe-type current to interpolate the baryon state and expect a better improvement.
\subsection{Light-cone QCD sum rules for the transition form factors}
The derivation of the heavy-light transition form factors is from the following correlation function:
\begin{equation}
T_\mu(P,q)=i \int d^4xe^{iq\cdot x}\langle 0|T\{j_{\Xi_c}(0)j_\mu^{w}(x)\}|\Sigma^+(P,s)\rangle,\label{weakcorrelator}
\end{equation}
where the weak interaction current is $j_\mu^w(x)=\bar c(x)\gamma_\mu(1-\gamma_5)s(x)$, and the interpolating current for $\Xi_c$ baryon is used as the Ioffe-type one:
\begin{equation}
j_{\Xi_c}=\epsilon_{ijk}[s^i C\gamma_5\gamma_\mu c^j]\gamma^\mu q^k.
\end{equation}
The coupling of this type of current with the baryon state is defined by the coupling constant $\lambda_{1c}$:
\begin{equation}
\langle 0\mid j_{\Xi_c} \mid \Xi_c(P')\rangle=\lambda_{1c}M_{\Xi_c}\Xi_c(P'),\label{Xicoup}
\end{equation}
and the transition form factors are defined as:
\begin{eqnarray}
\langle
\Xi_c(P')|j_\mu^w|\Xi(P)\rangle=\bar{\Xi_c}(P')[f_1\gamma_\mu-i\frac{f_2}{M}\sigma_{\mu\nu}q^\nu-(g_1\gamma_\mu+i\frac{g_2}{M}\sigma_{\mu\nu}q^\nu)\gamma_5]\Xi(P).
\label{weakff}
\end{eqnarray}

In order to get the sum rules, we need to express the correlation function (\ref{weakcorrelator}) both phenomenologically and theoretically. In the following we take $\Xi_c^+$ as an example. On the one hand, the hadronic expression is gotten by making use of definitions (\ref{Xicoup}) and (\ref{weakff}):
\begin{eqnarray}
z^\nu T_\nu&=&\frac{\lambda_{1c}M_{\Xi_c}}{M_{\Xi_c}^2-P'^2}\{2P\cdot zf_1(q^2)+\frac{2P\cdot z}{M_{\Xi_c}}\!\not\!{q_\perp}f_2(q^2)-[M
f_1(q^2)-\frac{q^2}{M_{\Xi_c}}f_2(q^2)]\!\not\!z\nonumber\\
&&+[f_1(q^2)+f_2(q^2)+\frac{M}{M_{\Xi_c}}]\!\not\!z\!\not\!q-2P\cdot z\gamma_5g_1(q^2)+\frac{2P\cdot
z}{M_{\Xi_c}}\!\not\!{q_\perp}\gamma_5g_2(q^2)\nonumber\\
&&-(Mg_1(q^2)+Mg_2(q^2)-\frac{q^2}{M_{\Xi_c}}g_2(q^2))\!\not\!z\gamma_5-[g_1(q^2)-g_2(q^2)\nonumber\\
&&+\frac{M}{M_{\Xi_c}}g_2(q^2)]\!\not\!z\!\not\!q\gamma_5\}\Xi(P).
\end{eqnarray}
It is similar to the discussions in the above section that there are more Lorentz structures than that needed. In the analysis, we choose terms proportional to $1$ ($\gamma_5$) and $\!\not\!q_\perp$ ($\!\not\!q_\perp\gamma_5$) to get the sum rules.

On the other hand, by contracting the heavy $c$ quark and using the distribution amplitudes presented in Ref. \cite{xidas}, we can get the theoretical expression of the correlation function (\ref{weakcorrelator}). Then matching the two expressions and making Borel transformation on both sides, we arrive at the final sum rules:

\begin{eqnarray}
2\lambda_{1c}M_{\Xi_c}f_1(q^2)&=&\int_{\alpha_{20}}^1d\alpha_2e^{-(s-M_{\Xi_c})/M_B^2}\Big\{M[2C_0-3C_0'-D_0](\alpha_2)+\frac{M}{\alpha_2}[C_1+4D_1\nonumber\\
&&-\frac{m_c}{M}E_0](\alpha_2)+\frac{M(m_c^2-q^2)}{\alpha_2^2M_B^2}[C_2+D_1](\alpha_2)+\frac{M^3}{M_B^2}[-C_2+2C_3\nonumber\\
&&-\frac{2}{\alpha_2}C_4](\alpha_2)-\frac{Mq^2}{\alpha_2M_B^2}D_1(\alpha_2)+\frac{2M^2m_c}{\alpha_2M_B^2}[E_1+E_2+\frac{1}{\alpha_2}E_3](\alpha_2)\nonumber\\
&&-\frac{2M^2m_c}{\alpha_3^2M_B^3}[Mm_cC_4+q^2E_3](\alpha_2)\Big\}\nonumber\\
&&+e^{-(s_0-M_{\Xi_c})/M_B^2}\frac{1}{\alpha_{20}^2M^2+m_c^2-q^2}\Big\{M(m_c^2-q^2)[C_1-D_1](\alpha_{20})\nonumber\\
&&-2M^3\alpha_{20}^2[C_2-C_3+\frac{1}{\alpha_{20}}C_4](\alpha_{20})-Mq^2D_1(\alpha_{20})+2M^2m_c\alpha_{20}\nonumber\\
&&\times[E_1+E_2+\frac{1}{\alpha_{20}}E_3](\alpha_{20})-\frac{2M^3m_c}{\alpha_{20}M_B^2}[Mm_cC_4-q^2E_3](\alpha_{20})\nonumber\\
&&+2\alpha_{20}^2\frac{d}{d\alpha_{20}}\frac{M^2m_c}{\alpha_{20}(\alpha_{20}^2M^2+m_c^2-q^2)}[Mm_cC_4-q^2E_3](\alpha_{20})\Big\}.\label{sumrulef1}
\end{eqnarray}
\begin{eqnarray}
2\lambda_{1c}\frac{M_{\Xi_c}}{M}f_2(q^2)&=&\int_{\alpha_{20}}^1d\alpha_2e^{-(s-M_{\Xi_c}^2)/M_B^2}\Big\{-\frac{1}{\alpha_2}[C_2-6D_0'](\alpha_2)+\frac{M^2}{\alpha_2M_B^2}[C_1+2C_2\nonumber\\
&&-2C_3+D_1+D_2](\alpha_2)+\frac{Mm_c}{\alpha_2^2M_B^2}[E_1-E_5](\alpha_2)-\frac{2M^3m_c}{\alpha_2^2M_B^4}E_3(\alpha_2)\Big\}\nonumber\\
&&+e^{-(s_0-M_{\Xi_c}^2)/M_B^2}\frac{1}{\alpha_{20}^2M^2+m_c^2-q^2}\Big\{\alpha_{20}M^2[C_1+2C_2-2C_3\nonumber\\
&&+D_1+D_2](\alpha_{20})+Mm_c[E_1-E_5](\alpha_{20})-\frac{2M^3m_c}{M_B^2}E_3(\alpha_{20})\nonumber\\
&&+2\alpha_{20}^2\frac{d}{d\alpha_{20}}\frac{M^3m_c}{\alpha_{20}^2M^2+m_c^2-q^2}E_3(\alpha_{20})\Big\}.\label{sumrulef2}
\end{eqnarray}
\begin{eqnarray}
-2\lambda_{1c}M_{\Xi_c}g_1(q^2)&=&\int_{\alpha_{20}}^1d\alpha_2e^{-(s-M_{\Xi_c})/M_B^2}\Big\{\frac{M}{\alpha_2}[\frac12C_1-\alpha_2C_0'+2\alpha_2D_0'+3\alpha_2D_0\nonumber\\
&&-3D_1-2\frac{m_c}{M}E_0](\alpha_2)+\frac{M(m_c^2+\alpha_2^2M^2+q^2)}{2\alpha_2^2M_B^2}C_1(\alpha_2)\nonumber\\
&&+\frac{M^3}{\alpha_2M_B^2}[\alpha_2[C_3+D_2-2D_3]+D_4+\frac{2m_c}{M}E_1-2\frac{m_c}{M}E_2\nonumber\\
&&+\frac{2m_c}{\alpha_2M}E_3](\alpha_2)+\frac{2M^2m_c}{\alpha_2^3M_B^4}[Mm_c\alpha_2D_4-q^2E_3-\alpha_2^2M^2E_4](\alpha_2)\Big\}\nonumber\\
&&+e^{-(s_0-M_{\Xi_c})/M_B^2}\frac{1}{\alpha_{20}^2M^2+m_c^2-q^2}\Big\{\frac{M(m_c^2+\alpha_{20}^2M^2+q^2)}{2}C_1(\alpha_{20})\nonumber\\
&&+\alpha_{20}M^3[\alpha_{20}[C_3+D_2-2D_3]+D_4+\frac{2m_c}{M}[E_1-E_2]\nonumber\\
&&+\frac{2m_c}{\alpha_{20}^2M}E_3](\alpha_{20})+\frac{1}{\alpha_{20}M_B^2}[2M^3m_c^2\alpha_{20}D_4-2M^2m_cq^2E_3\nonumber\\
&&-2M^4m_c\alpha_{20}^2E_4](\alpha_{20})+2\alpha_{20}^2\frac{d}{d\alpha_{20}}\frac{M^2m_c}{\alpha_{20}(\alpha_{20}^2M^2+m_c^2-q^2)}\nonumber\\
&&\times[Mm_c\alpha_{20}C_3-q^2D_3-\alpha_{20}^2M^2E_4](\alpha_{20})\Big\}.\label{sumruleg1}
\end{eqnarray}
\begin{eqnarray}
2\lambda_{1c}\frac{M_{\Xi_c}}{M}g_2(q^2)&=&\int_{\alpha_{20}}^1d\alpha_2e^{-(s-M_{\Xi_c}^2)/M_B^2}\Big\{\frac{1}{\alpha_2}[C_0+D_0'](\alpha_2)+\frac{M^2}{\alpha_2M_B^2}[C_1+C_2+D_1\nonumber\\
&&-2D_3-D_2](\alpha_2)+\frac{Mm_c}{\alpha_2^2M_B^2}[E_1-E_5](\alpha_2)-\frac{2M^3m_c}{\alpha_2^2M_B^4}[E_3+E_4](\alpha_2)\Big\}\nonumber\\
&&+e^{-(s_0-M_{\Xi_c}^2)/M_B^2}\frac{1}{\alpha_{20}^2M^2+m_c^2-q^2}\Big\{\alpha_{20}M^2[C_1+C_3+D_1-2D_3\nonumber\\
&&-D_2](\alpha_{20})+Mm_c[E_1-E_5](\alpha_{20})-\frac{2M^3m_c}{M_B^2}[E_3+E_4](\alpha_{20})\nonumber\\
&&+2\alpha_{20}^2\frac{d}{d\alpha_{20}}\frac{M^3m_c}{\alpha_{20}^2M^2+m_c^2-q^2}[E_3+E_4](\alpha_{20})\Big\}.\label{sumruleg2}
\end{eqnarray}

where the following expressions are used for convenience:
\begin{eqnarray}
D_3(\alpha_2)&=&(-\widetilde A_1-\widetilde A_3+\widetilde A_5)(\alpha_2),\nonumber\\
D_4(\alpha_2)&=&(\widetilde{\widetilde A}_1-\widetilde{\widetilde A}_2+\widetilde{\widetilde A}_3+\widetilde{\widetilde A}_4-\widetilde{\widetilde A}_5+\widetilde{\widetilde A}_6)(\alpha_2).
\end{eqnarray}
The other expressions are used the same as that appearing in Section \ref{sec2}.
\subsection{Numerical analysis}
Before the numerical analysis on the semileptonic process, the nonperturbative parameter $\lambda_{1c}$ is determined in the QCD sum rules. Following the standard procedure, we get the final result:
\begin{eqnarray}
(4\pi)^4\lambda_{1c}^2M_{\Xi_c}^2e^{-M_{\Xi_c}^2/M_B^2}&=&\int_{(m_c+m_s)^2}^{s_0}e^{-s/M_B^2}ds\Big\{\frac43s^2[1-8x+8x^3-x^4-12x^2\ln x]\nonumber\\
&&+4m_cm_ss[-2-3x+6x^2-x^3-6x\ln x]\nonumber\\
&&+2m_ca_s(1-x)^2-(2m_c+m_s)m_0^2a_s\frac{x}{s}-\frac{m_sa_s}{2}(1-x^2)\nonumber\\
&&-\frac{b}{6}(1-x)(5-x)-\frac{2b}{3}m_cm_s\frac{1}{s}\frac{1-x^2}{x}\Big\},
\end{eqnarray}
where $x=m_c^2/s$, and the other parameters are used as the standard values in the QCD sum rule: $a=-(2\pi)^2\langle\bar uu\rangle=0.55\; \mbox{GeV}^{3}$,
$b=(2\pi)^2\langle\alpha_sG^2/\pi\rangle=0.47\; \mbox{GeV}^{4}$, $a_s=-(2\pi)^2\langle\bar ss\rangle=0.8a$, $\langle\bar ug_c\sigma\cdot Gu\rangle=0.8\langle\bar
uu\rangle$. As comments in Ref. \cite{xidas}, the heavy baryon mass is set to be $M_{\Xi_c}=2.471\,\mbox{GeV}^2$. It is noted that in the paper the uncertainties originating from some input parameters, such as the baryon masses, the strange quark mass and the condensates, have not been
considered due to the fact that the sum rule method itself has about $20$ percent uncertainties, which makes it less significant to take into account
the errors of the input parameters. With the threshold $s_0=7.8\,\mbox{GeV}^2$ and the Borel parameter in the working window $1.1\,\mbox{GeV}^2\le M_B^2\le 1.6\,\mbox{GeV}^2$, we get the result $\lambda_{1c}=-(0.011\pm0.001)\,\mbox{GeV}^2$.

\begin{figure}
\begin{minipage}{7cm}
\epsfxsize=6cm \centerline{\epsffile{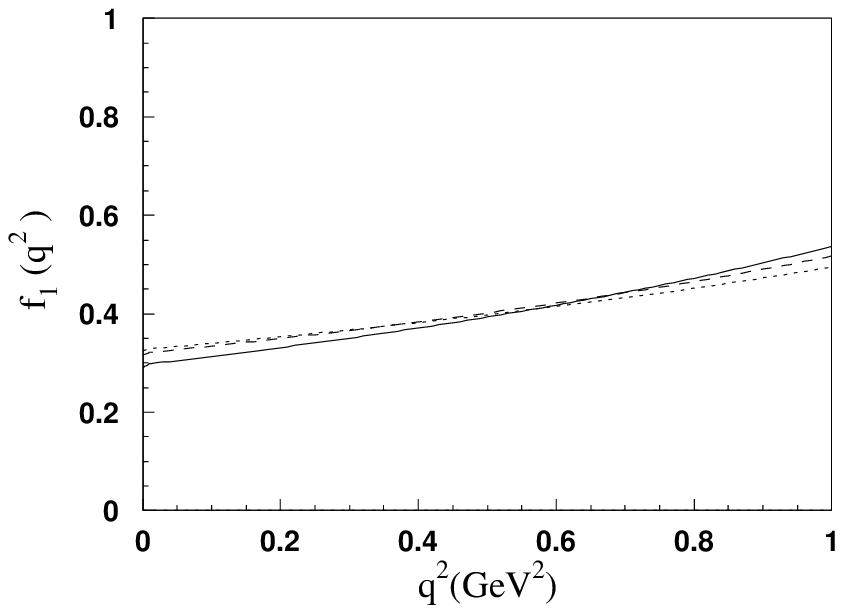}}
\end{minipage}
\hfill
\begin{minipage}{7cm}
\epsfxsize=6cm \centerline{\epsffile{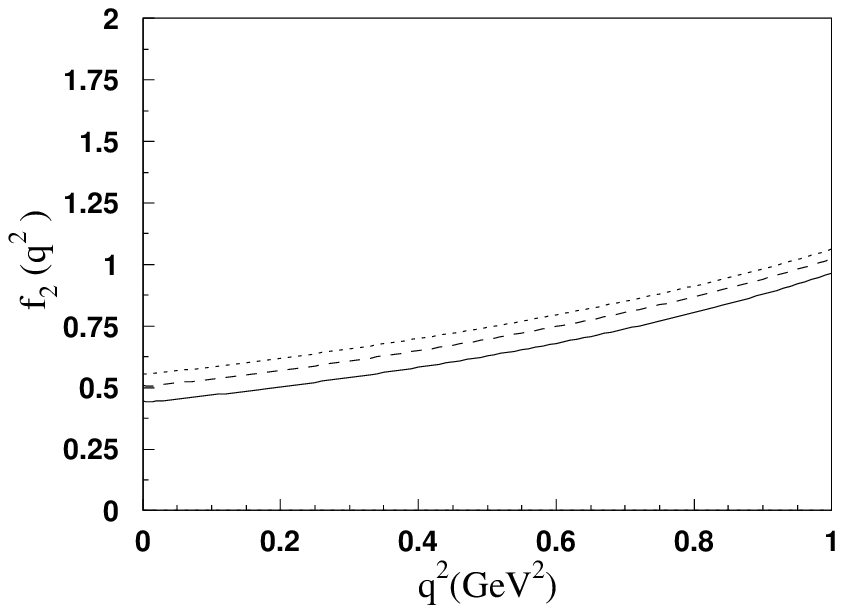}}
\end{minipage}
\hfill
\begin{minipage}{7cm}
\epsfxsize=6cm \centerline{\epsffile{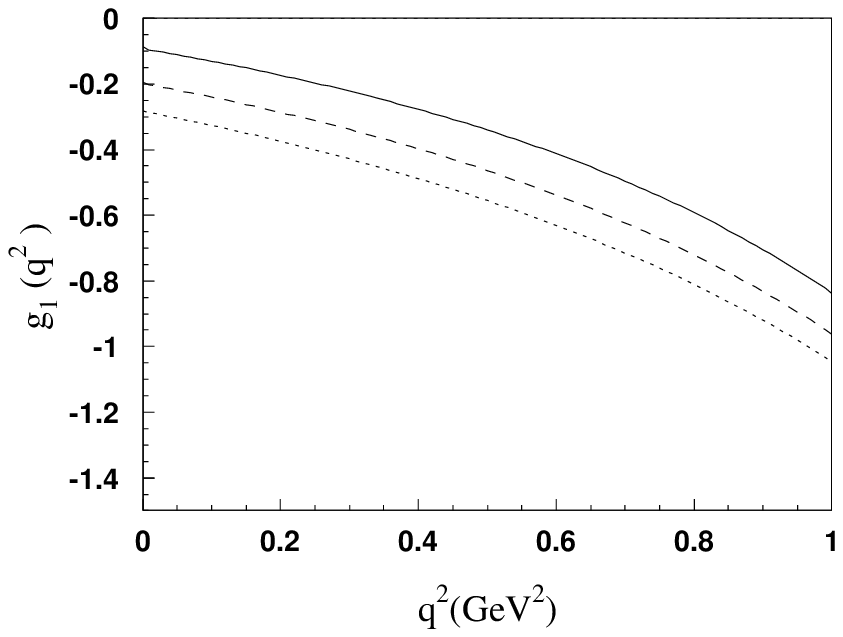}}
\end{minipage}
\hfill
\begin{minipage}{7cm}
\epsfxsize=6cm \centerline{\epsffile{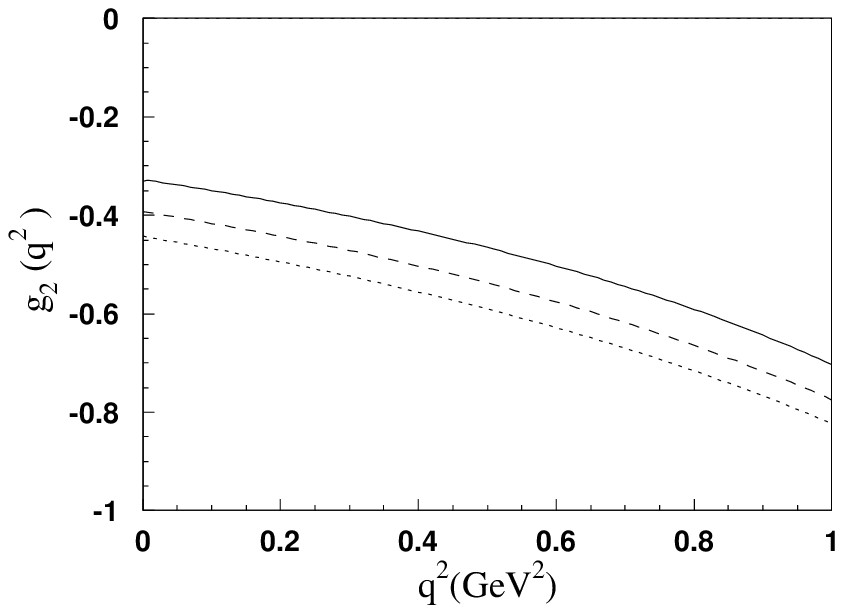}}
\end{minipage}
\caption{}\label{fig3}
\end{figure}

For the numerical analysis of the transition form factors, the results show that the sum rules are working well in the Borel working region $7\,\mbox{GeV}^2\le
M_B^2\le 9\,\mbox{GeV}^2$ with the threshold $7\,\mbox{GeV}^2\le s_0\le 9\,\mbox{GeV}^2$. In Fig. \ref{fig3} we present the dependence of the form factors on the momentum transfer $q^2$.

It is the same as what has been discussed in Refs. \cite{lamc,xidas} that we can only give the form factors in the light-cone sum rule allowed region $0\le q^2\le 1\,\mbox{GeV}^2$, while the thorough understanding of the process requires us to know information of the whole dynamical region, e.g. $0\le q^2\le (M_{\Xi_c}-M_\Xi)^2$. For this reason, we fit the form factors by the three-parameter dipole formula:
\begin{equation}
f(q^2)=\frac{f(0)}{1+a(q^2/{M_{\Xi_c}}^2)+b(q^2/{M_{\Xi_c}}^2)^2}.\label{weakformulafit}
\end{equation}
The fits are shown in Table \ref{table2}.

\begin{figure}
\begin{minipage}{7cm}
\epsfxsize=7cm \centerline{\epsffile{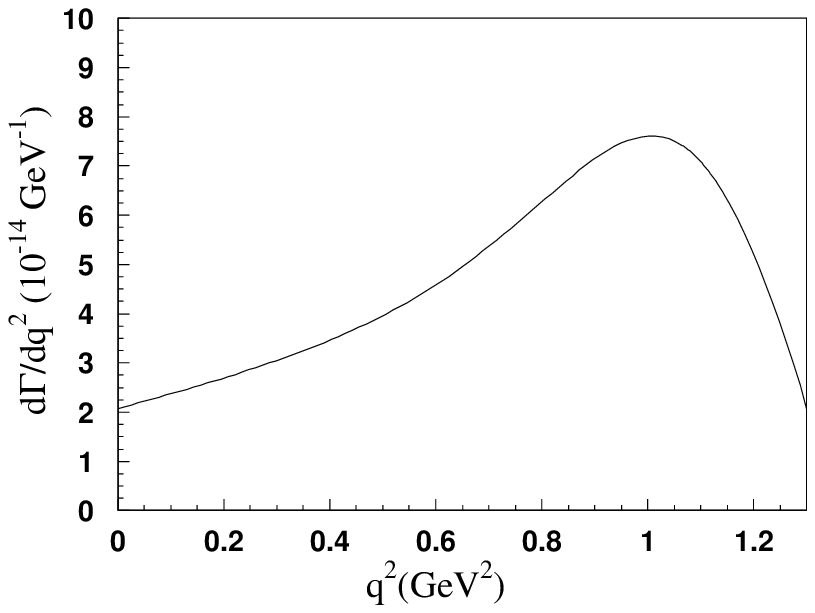}}
\end{minipage}
\caption{}\label{fig4}
\end{figure}

Then we extrapolate the fit formula to the whole physical region and get the differential decay rate by the same formula in Refs. \cite{lamc,xidas}, which is shown in Fig. \ref{fig4}.

The total decay width is estimated from the differential decay rate by integrating out the momentum transfer $q^2$ in the whole dynamical area. The final result is $\Gamma(\Xi_c\rightarrow \Xi e^+\nu_e)=(6.17^{+2.24}_{-2.48})\times10^{-14}\;\mbox{GeV}$. The error in the numerical result comes from the different choice of the threshold and the Borel parameter varying in their working region. The decay mode has been observed for several years, but the absolute branching fraction is still not measured yet\cite{PDG}. For a comparison, we turn to experimental data from Ref. \cite{Alexander}, where the authors have given the relative branching ratios of the process: $B(\Xi_c^+\rightarrow \Xi^-\pi^+\pi^+)/B(\Xi_c^+\rightarrow\Xi^0 e^+\nu_e)$ and $B(\Xi_c^0\rightarrow \Xi^-\pi^+)/B(\Xi_c^0\rightarrow\Xi^- e^+\nu_e)$. With the up bounds of the channels $B(\Xi_c^+\rightarrow \Xi^-\pi^+\pi^+)\leq2.1\times10^{-2}$ and $B(\Xi_c^0\rightarrow \Xi^-\pi^+)\leq4.3\times10^{-3}$, our estimations are given as $B(\Xi_c^+\rightarrow\Xi^0 e^+\nu_e)/B(\Xi_c^+\rightarrow \Xi^-\pi^+\pi^+)=2.0^{+0.7}_{-0.8}$ and $B(\Xi_c^0\rightarrow\Xi^- e^+\nu_e)/B(\Xi_c^0\rightarrow \Xi^-\pi^+)=2.4^{+0.9}_{-1.0}$, which is more reliable in comparison with the results from the CZ-type current. The comparison is shown in Table \ref{table3}.

In summary, we investigate the electromagnetic form factors of the $\Xi$ baryon and study the semileptonic decay process of the $\Xi_c$ baryon. The calculation is carried out in the light cone QCD sum rule approach with the Ioffe-type interpolating currents. The magnetic moments of the baryons are estimated by the dipole formula fits of the magnetic form factor. Our estimates are $\mu_{\Xi^0}=-(1.75\pm0.21)\mu_N$ and $\mu_{\Xi^-}=-(1.01\pm0.11)\mu_N$, which improve the results in comparison with the case where the CZ-type current is used. The decay width of the semileptonic decay processes $\Xi_c\rightarrow \Xi e^+\nu_e$ is estimated to be $\Gamma(\Xi_c\rightarrow \Xi e^+\nu_e)=(6.17^{+2.24}_{-2.48})\times10^{-14}\;\mbox{GeV}$, which is in accordance with the existing experiment. By comparing our estimates with the experimental data, it is found that the adoption of the Ioffe-type current may give more reliable results.

\newpage

\begin{center}
\begin{minipage}{12cm}
\baselineskip 24pt {\sf Fig. 1.}{\quad Dependence of the $\Xi^0$ baryon magnetic (a) and electric (b) form factors on the momentum transfer. The dashed, solid, and dotted lines correspond to the threshold $s_0=2.8\,,2.9\,,3.0\;\mbox{GeV}^2$, respectively.}
\end{minipage}
\end{center}

\begin{center}
\begin{minipage}{12cm}
\baselineskip 24pt {\sf Fig. 2.}{\quad Dependence of the $\Xi^-$ baryon magnetic (a) and electric (b) form factors on the momentum transfer. The three lines correspond to the threshold $s_0=2.8\,,2.9\,,3.0\;\mbox{GeV}^2$ from the top down.}
\end{minipage}
\end{center}

\begin{center}
\begin{minipage}{12cm}
\baselineskip 24pt {\sf Fig. 3.}{\quad Dependence of the transition form factors of the process $\Xi_c\rightarrow\Xi$ on the momentum transfer $q^2$ with $M_B^2=8\,\mbox{GeV}^2$. The solid, dashed and dotted lines correspond to the threshold $s_0=7,\,8,\,9\,\mbox{GeV}^2$, respectively.}
\end{minipage}
\end{center}

\begin{center}
\begin{minipage}{12cm}
\baselineskip 24pt {\sf Fig. 4.}{\quad The differential decay rate of the semileptonic process $\Xi_c\rightarrow \Xi e^+\nu_e$.}
\end{minipage}
\end{center}

\newpage

\begin{table}
\renewcommand{\arraystretch}{1.1}
\caption{The magnetic moments of the baryons from various interpolating currents.}
\begin{center}
\begin{tabular}{|c|c|c|}
\hline
$\mu$($\mu_N$)&$\Xi^0$&$\Xi^-$\\
\hline
CZ current& $-1.92\pm0.34$ & $-1.19\pm0.03$  \\
\hline
Ioffe current& $-1.75\pm0.21$ & $-1.01\pm0.11$ \\
\hline
PDG& $-1.250\pm0.014$ & $-0.6507\pm0.0025$ \\
\hline
\end{tabular}
\end{center} \label{table1}
\end{table}

\clearpage

\newpage

\begin{table}[h]
\caption{Three-parameter dipole formula fits of the weak transition form factors for the process $\Xi_c\rightarrow\Xi e^+\nu_e$.}
\begin{center}
\begin{tabular}
{|c|c|c|c|}
\hline  & $f_i(0)$ & $a_1$ & $a_2$  \\
\hline  $f_1$ & $0.32$ & $-2.64$ & $1.31$  \\
\hline  $f_2$ & $0.50$ & $-3.65$ & $3.34$  \\
\hline  $g_1$ & $-0.22$ & $-7.93$ & $19.85$  \\
\hline  $g_2$ & $0.39$ & $-3.77$ & $3.45$  \\
\hline
\end{tabular}
\end{center}\label{table2}
\end{table}

\clearpage

\newpage

\begin{table}[htb]
\caption{The prediction of the process $\Xi_c\rightarrow\Xi e^+\nu_e$ with two kinds of interpolating currents.}
\begin{center}
\begin{tabular}{|c|c|c|c|c|c|c|}
\hline &\multicolumn{3}{|c|}{$\Gamma(\times10^{-14})\mbox{GeV}$} &\multicolumn{3}{|c|}{$Br(\%)$}\\
\hline
& CZ-tpye & Ioffe-type& PDG& CZ-tpye & Ioffe-type & PDG\\
\hline
$\Xi_c^+\rightarrow\Xi^0 e^+\nu_e$&$8.73$ &$6.17$&$-$&$2.7$ &$2.0$&$2.3$ \\
\cline{1-1}\cline{2-2}\cline{3-3}\cline{5-5}\cline{6-6}\cline{7-7}
$\Xi_c^0\rightarrow\Xi^-e^+\nu_e$&$8.73$&$6.17$&&$3.4$&$2.4$&$3.1$\\
\hline
\end{tabular}
\end{center}\label{table3}
\end{table}
\clearpage

\newpage

\clearpage

\newpage

\clearpage

\newpage

\clearpage

\newpage

\end{document}